\documentclass{sig-alternate}

\usepackage{enumitem,url,color,graphicx,xspace,amsmath,listings,times}
\usepackage{caption,balance,mathtools,subcaption}
\usepackage[square,sort,comma,numbers]{natbib}
\usepackage[table]{xcolor}
\usepackage{todonotes}

\begin{document}

\title{Adding Real-time Capabilities to a SML Compiler}

\author{
Muyuan Li,
Daniel E McArdle,
Jeffrey C Murphy,
Bhargav Shivkumar,
Lukasz Ziarek\\
\affaddr{SUNY Buffalo}\\
\email{\{muyuanli,demcardl,jcmurphy,bhargavs,lziarek\}@buffalo.edu}
}

\maketitle

\begin{abstract}
There has been much recent interest in adopting functional and reactive programming for use in real-time system design.
Moving toward a more declarative methodology for developing real-time systems purports to improve the fidelity of
software. To study the benefits of functional and reactive programming for real-time systems, real-time aware functional
compilers and language runtimes are required. 
In this paper we examine the necessary changes to a modern Standard ML compiler, MLton, to provide basic support for real-time
execution.  We detail our current progress in modifying MLton with a threading model that supports priorities, a chunked object
model to support real-time garbage collection, and low level modification to execute on top of a real-time operating system. We
present preliminary numbers and our work in progress prototype, which is able to boot ML programs compiled with MLton on x86 machines.

\end{abstract}

\section{Introduction}
\label{sec:intro}
Recent work on functional reactive programming has once again spurred interest in examining the usage
of functional and declarative languages for main stream real-time system design and programming~\cite{Taha:2001:DFP:646787.703875,wan-phd,Jiang:2014:MST:2760441.2760854,Belwal:2011:FIT:2117692.2119052}.
Functional programming, much like the mantra of real-time Java, provides a type-safe vehicle for
real-time system implementation that by nature of the language structure itself prevents common errors and bugs,
like buffer under/over flow and null pointer dereference, from being expressed. Programmers can thus
produce higher fidelity code with lower programmer effort.   Additionally, functional programming languages
are typically easier to analyze statically than their object oriented counter parts, and significantly easier
than C. As such, they purport to reduce time and effort from a validation and verification perspective.

As the research community explores new programming models that leverage the benefits of functional
and reactive programming, there is an increasing need to examine the addition of real-time capabilities
to functional language runtimes. Additionally, classic optimization strategies that most compilers for functional
programming languages leverage to achieve performance need to be revisited and adapted for
{\em predictable} execution. 

In this paper we present a work in progress system that examines what mechanisms must be in the language runtime to facilitate the development of functional, real-time systems. 
Specifically we investigate what changes need to be made to MLton~\cite{mlton}, a modern SML compiler, in order to add real-time capabilities.
This includes an overhaul of the threading system to support fixed priority based scheduling, a new
chunked object model for predictable allocation and non moving real-time garbage collection, as well as bindings for
a real-time OS for embedded deployment. We leverage our previous experience with Multi-MLton~\cite{Sivaramakrishnan:2010:LAU:1708046.1708059,JFP:9547587} and the Fiji real-time virtual machine~\cite{pizlo:eurosys:2010}
in guiding our modifications to MLton.  Our changes sit below the MLton library level, providing building blocks to explore new programming models.
We present preliminary performance numbers, indicating the viability of our prototype.
Our prototype supports embedded execution on x86 architectures and is publicly available for download at: \url{https://github.com/UBMLtonGroup}.

%
%
%

\section{MLton Overview}
\label{sec:system}

MLton~\cite{mlton,Ziarek:2008:FTS:1466762.1466777} is a whole-program optimizing compiler for SML 97. 
MLton's leverages whole-program optimization using a simply-typed first-order intermediate language.
There are numerous issues that arise when translating SML into a simply-typed IL. First, how does one represent SML modules and
functors, since these typically require much more complicated type systems?
MLton's answer: defunctorize the program~\cite{Reynolds:1998:DIH:609145.609184}. This transformation turns an SML program with modules into an
equivalent one without modules by duplicating each functor at every application and eliminating structures by renaming
variables. Second, how does one represent SML's polymorphic types and polymorphic functions?
MLton's answer: monomorphise the program~\cite{Tolmach:1998:MAS:969592.969596}. This transformation eliminates polymorphism from
an SML program by duplicating each polymorphic datatype and function at every type at which it is instantiated.
Third, how does one represent SML's higher-order functions? MLton's answer: defunctionalize the
program. This transformation replaces higher-order functions with data
structures to represent them and first-order functions to apply them; the resulting IL is Static Single Assignment form.
Because each of the above transformations requires matching a functor, function definition, or type definition with
all possible uses, MLton must be a whole-program compiler. MLton's whole-program compilation strategy has a number of
implications. Most importantly, MLton's use of defunctorization means that the placement of code in modules has no effect
on performance. In fact, it has no effect on the generated code whatsoever. Modules are purely for the benefit
of the programmer in structuring code. Also, because MLton duplicates functors at each use, no run-time penalty
is incurred for abstracting a module into a functor. The benefits of monomorphisation are similar. Thus,
with MLton, a programmer does not suffer the time and space penalties from an extra level of indirection
in a list of doubles just because the compiler needs a uniform representation of lists.

We believe that the MLton approach is beneficial for real-time programming as it yields very efficient code, both in time and space.
MLton's whole program optimization strategy provides us with precise low-level information about object layouts and sizes, which can be leveraged
to optimize existing real-time garbage collection approaches proposed for real-time Java.
MLton is easily adaptable to an embedded workflow, by having MLton emit ANSI-C code and then
cross compiling using specialized C cross-compilers provided by most real-time OS vendors.  Unfortunately, MLton was
not designed with real-time applications in mind. Specifically, MLton's threading model and GCs are neither
priority aware nor predictable. In the following sub-sections we discuss MLton's internals and their implication on
real-time execution.

%

\subsection{Threading}

MLton provides a concurrent, but not parallel, threading model. As such, MLton created threads are green threads that are multiplexed over a single OS level thread. MLton's thread API is well suited for implementing user defined schedulers, including preemptive and cooperative threading models as well as Concurrent ML~\cite{cml}.
A thread in MLton is a lightweight data structure that represents a paused computation. Threads contain the currently saved execution state of the program, namely the call stack. When a thread is paused, a copy of its current stack is saved and when it is switched to, the stack is restored. MLton also provides a ready queue from which the next runnable thread is accessed by the scheduler. This is a regular FIFO queue with no notion of priority, however the structure is {\em implicit}, relying on continuation chaining and is embedded in the thread switching code. Threading libraries build on top of the MLton thread primitives typically leverage a thread queue data structure ({\em e.g.} CML).

One of the main prerequisites of a real-time programming language is a threading model consisting of an analyzable scheduling mechanism as well as a scheduling algorithm.
MLton's implementation of threading does involve the underlying operating system in any way. The absence of direct relation between a MLton green thread and an OS level thread, results in the operating system seeing only one MLton thread and schedules it for execution along with other non-MLton threads in the system.  All MLton green threads are considered equal and the existing model preempts threads either after a fixed interval of time (preemptive) or when the currently executing thread decides to yield (cooperative). The absence of the notion of priority is quite crucial to the use of MLton for real-time systems as the criticality of computations are ignored. Such a case can be explicitly observed when any MLton thread makes a blocking IO call. Since all the green threads are mapped onto a single OS thread, this blocks all other MLton green threads as well until the OS finishes the execution of the IO operation and the blocking green thread can be descheduled to allow another thread to execute.

MLton's execution model places the SML call stack on the heap during the compilation, when it maps SML functions to C code.
MLton requires the intervention of the GC when a thread has to grow its stack during execution. The GC clears up space on the heap and then allocates a bigger stack and copies the existing stack into the new location. This process requires the execution of the current thread to pause until the GC exits, thus introducing a point of non deterministic overhead that is undesirable in a real-time system. These properties of MLton make it unsuitable for use as a real-time system out of the box.

\subsection{Garbage Collection}
\begin{figure}[htbp]
  \centering
  \includegraphics[width=0.75\columnwidth]{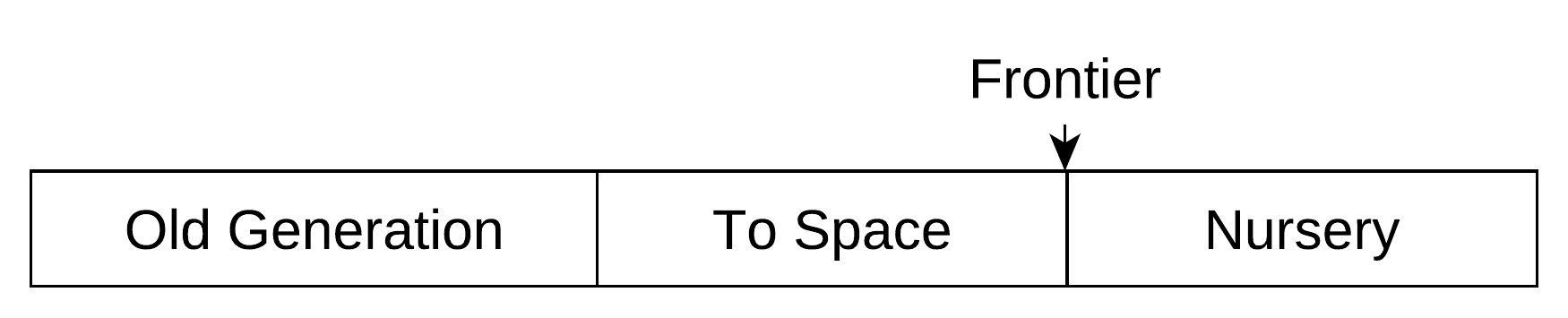}
  \caption{Heap layout in MLton}
  \label{fig:mlton-heap}
\end{figure}

MLton adopts a hybrid model of Cheney Copy GC and Mark-Compact. The
garbage collector dynamically switches between the two schemes back
and forth based on runtime memory utilization. Its heap layout is
depicted in Fig.~\ref{fig:mlton-heap}.

The default GC scheme is Cheney Copy GC, in which the new objects are
allocated in Nursery. When there is insufficient space, a minor Cheney
Copy is used to move objects from the nursery to "to space". If a minor GC
fails to collect enough heap space for the new object, a major Cheney
Copy GC is used, in which a secondary heap is allocated and the GC
will attempt to copy objects from current heap to secondary heap.
When the total memory usage exceeds 50\%, the GC falls back to a
2-generation Mark-Compact GC where the objects are moved to the old
generation if the live ratio is low.

MLton has 4 types of objects: normal object, arrays, weak object and
stack. These objects are treated differently in GC allocation. Arrays
and stacks are typically allocated in Old Generation, since they tend
to persist over the lifetime of the program. MLton makes bump-pointer
allocation at the frontier for normal and weak objects. However, weak
objects are not common -- they are typically created by programmers
explicitly via primitive calls.

\begin{figure}[htbp]
  \centering
  \includegraphics[width=0.8\columnwidth]{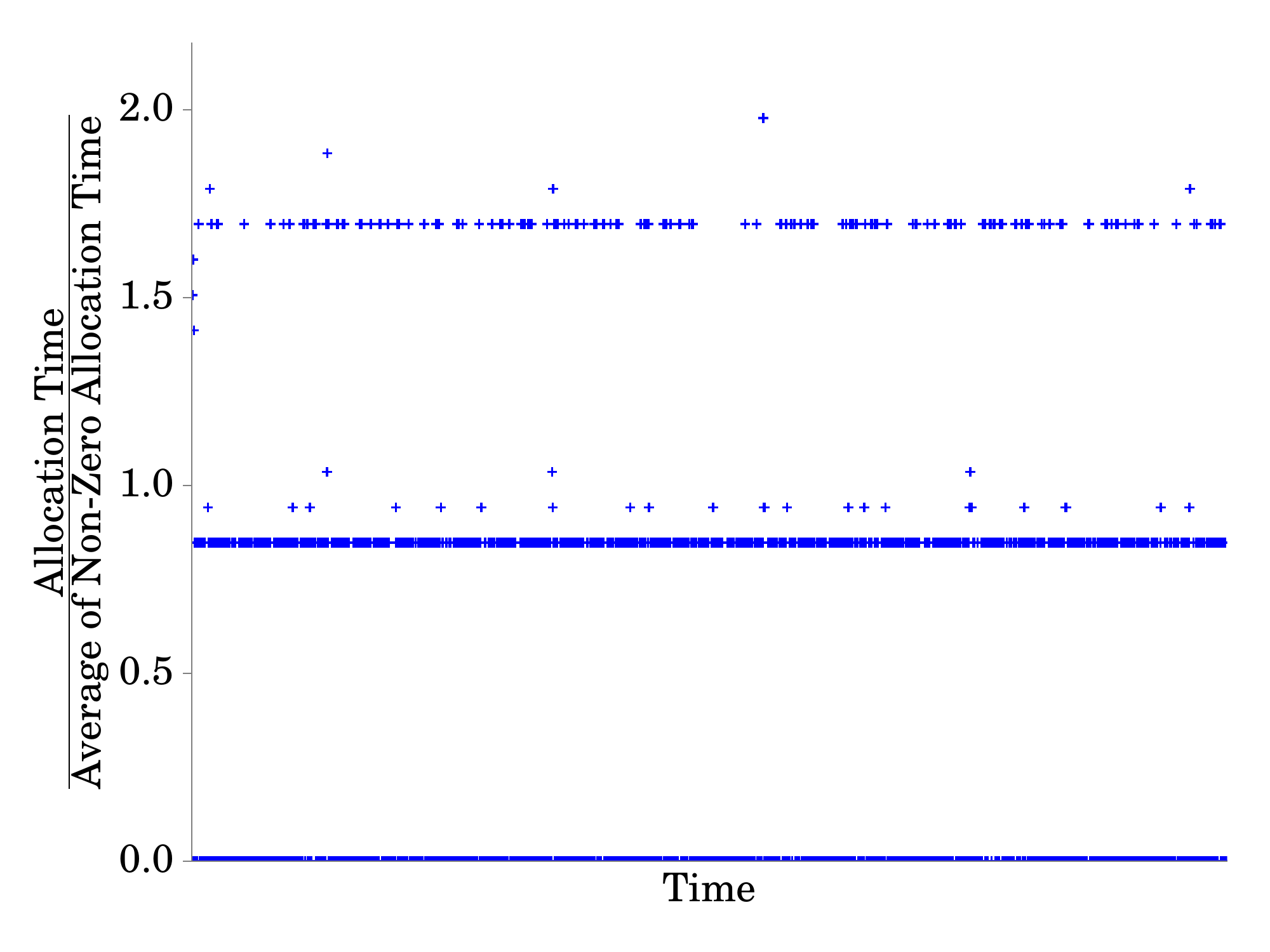}
  \caption{Unpredictable object allocation time in MLton}
  \label{fig:mlton-unpredict}
\end{figure}

MLton's GC scheme moves objects to achieve compaction, which can be a
source of unpredictability. Fig.~\ref{fig:mlton-unpredict} depicts
such unpredictability. We preform a micro benchmark on MLton by
allocating {\tt int array option} objects, in which we randomly choose
between allocating {\tt NONE} or {\tt SOME array} of 10 million elements. MLton first tries to
allocate object by following a Cheney Copy scheme. Then it compacts
the heap by copying the object to To Space, which introduces
non-predictable behavior. In this benchmark, the time to move objects
randomly goes up to 2x with no obvious pattern.

\label{sec:impl}

\section{Threading Model}

\begin{figure}[htbp]
  \centering
  \includegraphics[width=.9\columnwidth]{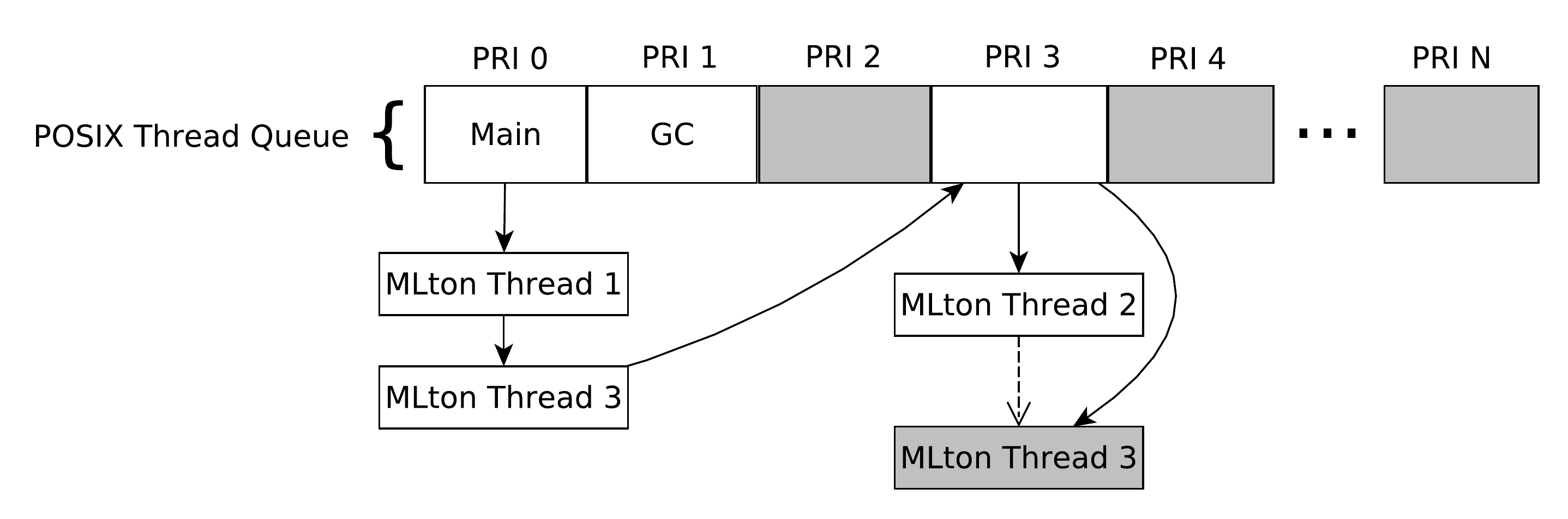}
  \caption{Priority-based Thread Queue}
  \label{fig:priQ}
\end{figure}

 In order to adapt MLton for use in real-time systems, we propose exposing the
POSIX threading API within MLton.  This will allow us to
propagate priority information from the ML thread tracking structure
to the RTOS and to leverage the RTOS scheduler.
Additionally, this model allows for the use of the established MLton
green threading models within each POSIX thread providing
a many-to-one mapping of green-to-OS threads. This model
also allows for the grouping of threads by activity,
for example IO versus computation, which in turn helps
isolate long-blocking activities such as IO into their own
threads. Fig.~\ref{fig:priQ} shows threads being isolated by priority as well as migrating between priorities,
but alternate models bucketing threads by activity
are also possible.

Our base scheduling mechanism is a fixed priority scheduler,
mapping one MLton thread of a given priority
to a single RTOS thread of the same priority. The maximum
number of threads is specified by the programer in a configuration parameter, which
allows for static preallocation of threads.
We note that this mirrors the approach taken by many RTOSes. We are currently
investigating a tiered scheduling mechanism, which allows multiple MLton threads
of the same priority to execute on top of a single RTOS thread with user defined
schedulers to dictate their scheduling policy. In order to handle priority inversion,we are looking into the use of Priority Inheritance Protocol locks~\cite{Sha:1990:PIP:102822.626613}.

In order to facilitate the use of OS level threads, several
things must change within MLton. For example, the GC must
first be moved to a separate thread and made concurrent. For
slack based RTGC support this also includes making the GC the lowest priority
thread in the system and supporting a concurrent GC that is interruptible at
any point~\cite{fijiGC}.  Secondly, the MLton model of a thread requesting GC
needs to be changed so that computation is never stalled on GC.


\section{Hybrid Fragmenting GC}
\label{sec:hybrid-gc}
The design of a real-time garbage collector should ensure
predictability. To eliminate GC pauses induced by
defragmentation and compacting the heap, we make sure that objects are
allocated as fixed-size chunks so that objects will never need to be
moved for defragmentation through the use of a hybrid fragmenting GC~\cite{fijiGC}.
 Small (normal and weak) objects,
arrays and stacks are allocated on 3 separate regions of the heap. We
maintain the array region and the normal object region by 2 free
lists, whereas the stacks are allocated in bump-pointer manner.

Normal and weak objects are represented as linked lists. Since object
sizes in MLton are typically small, we achieve constant access time when
allocating these objects.
Arrays are represented as trees, in which each node is
fixed-size. Internal nodes have a large number of branches (32 in our
implementation), which keeps access time $log_{32}(n)$ and is close to
constant. MLton constantly allocates small sized arrays and even zero
sized arrays. We represent such arrays as a single leaf to
eliminate the overhead of finding the immediate child of the root.
During collection, the GC first marks all fixed-size
chunks that are currently live. Then it sweeps the heap and returns
all unmarked chunks to the two free lists.

{\textbf{Heap Layout:}}
In MLton, the size of normal objects, arrays and stacks vary significantly. To
minimize the overhead induced by object chunking, we partition the
heap into three regions for normal / weak objects, arrays and stacks.

{\textbf{Object Layout:}
MLton tries to pack small objects into larger ones. In our empirical
study, most MLton objects are around 24 bytes. We choose 32
bytes as the chunk payload that carries MLton object along with
extra 12 bytes overhead associated with chunk management. Objects that are
larger than 32 bytes are split into multiple chunks. In our current
implementation, we limit object to two chunks each since we haven't
noticed objects that are greater than 64 bytes. The object layout is
depicted in Fig.~\ref{fig:chunked-object}.

\begin{figure}[htbp]
  \centering
  \includegraphics[width=.8\columnwidth]{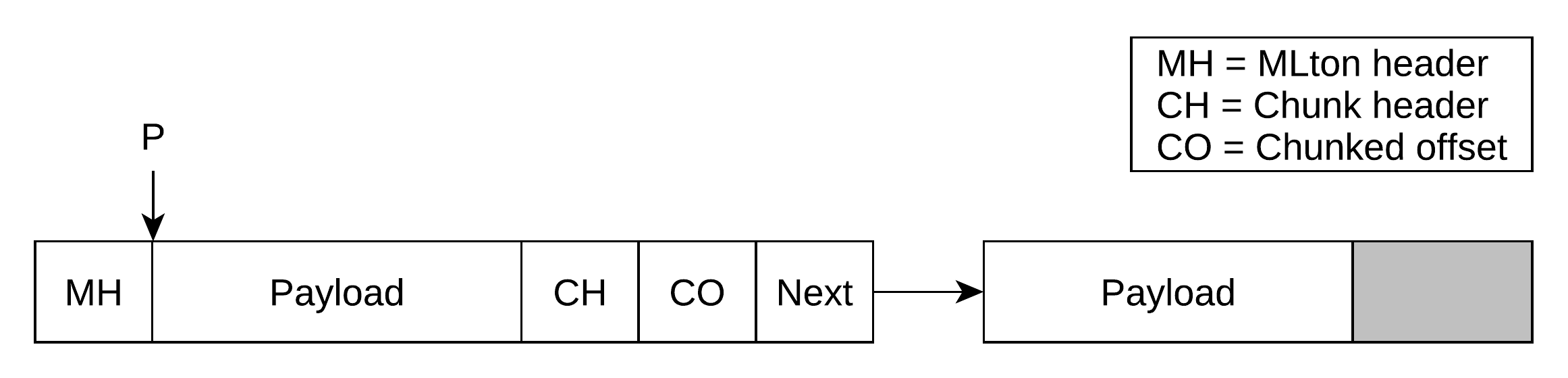}
  \caption{Chunked object layout}
  \label{fig:chunked-object}
\end{figure}

When an object fits into one chunk, the object field is calculated by:

\begin{lstlisting}[basicstyle=\footnotesize\ttfamily, breaklines=true]
  objectField(p, offset) = p - HEADER_SIZE + offset
\end{lstlisting}

\noindent If an object is chunked, the {\tt CO} field records the smallest
offset that causes the chunk. The object field is then retrieved by:

\begin{lstlisting}[basicstyle=\footnotesize\ttfamily, breaklines=true]
  objectField(p, offset) = (p - HEADER_SIZE)->next + offset - CO
\end{lstlisting}

Arrays are represented as trees. In MLton, arrays are typically passed
around using a pointer to its payload. The header and length of an array
are retrieved by subtracting the header size and array length size
from current pointer. We stick to this representation as much as
possible. Array nodes are represented in
Fig.~\ref{fig:chunked-array}. Internal nodes carry 32 pointers to
their children. We pass an array around via a pointer to its first leaf. A
root pointer and a next pointer is embedded in the leaf node. The leaf
pointer connects all leaves that actually carry payloads for
potential linear traversal optimization. For
an array that is 128 bytes or less, we can fit it into 1 leaf chunk. For
arrays that span multiple chunks, we construct trees. When accessing
an element of an array, we first follow the root pointer to retrieve
the root node and then access the array in a top down manner,
in which we determine the branch in
current node by {\tt index \% CO}, then we follow the branch to an
alternative internal node. The process is repeated until we finally
arrive at a leaf.

\begin{figure*}[htbp]
  \centering
  \includegraphics[width=.7\textwidth]{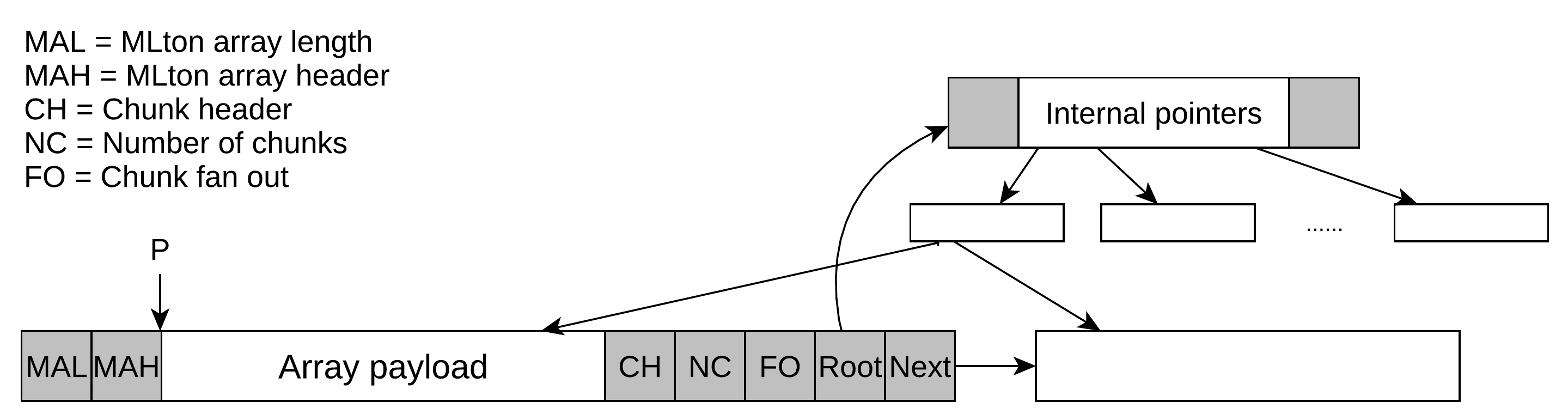}
  \caption{Chunked array layout}
  \label{fig:chunked-array}
\end{figure*}

In our current implementation, we leave the stack allocated by bump
pointer. The rationale is that, stacks never need to be moved in
our implementation. Stacks grow and shrink by pushing and popping
stack frames which results in chunks being added and removed as
necessary. They are self-managed and do not need GC intervention
to grow as they do in the MLton GC model.

Flattening refers to the multiple optimization passes in MLton that reduces
the overhead for accessing nested objects. Flattening in MLton changes
our array access scheme. The following code creates an array of tuples of
{\tt int} and {\tt bool} shown in Fig. \ref{fig:flattening-array}.
Then it reads the second part of the 4$^{\mbox{th}}$ element.

\begin{lstlisting}[basicstyle=\footnotesize\ttfamily, breaklines=true]
val arr = Array.tabulate (5, fn i => (i, true))
val a = #2 (Array.sub (arr, 4))
\end{lstlisting}

\begin{figure}[htbp]
  \centering
  \includegraphics[width=\columnwidth]{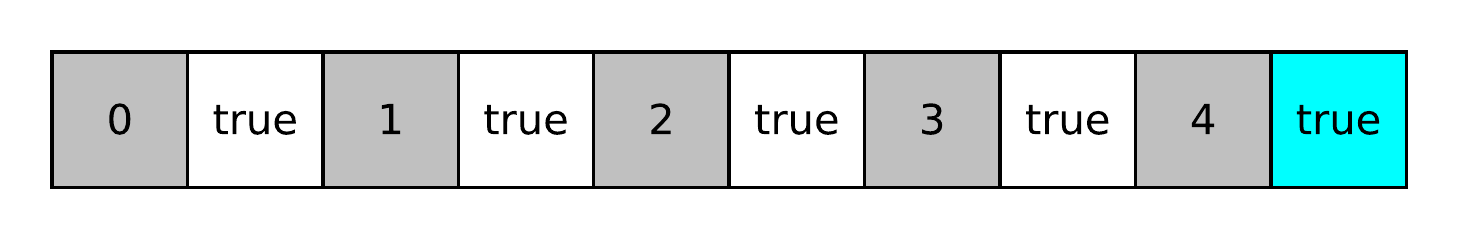}
  \caption{Accessing array of tuples}
  \label{fig:flattening-array}
\end{figure}

Without flattening, the read is translated into:

\begin{lstlisting}[basicstyle=\footnotesize\ttfamily, breaklines=true]
tup = arrayOffset(arr, elemSize=8, index=4)
a = objectOffset(tup, elemSize=4, index=1)
\end{lstlisting}

\noindent and with flattening, the read consists only a single element:

\begin{lstlisting}[basicstyle=\footnotesize\ttfamily, breaklines=true]
a = arrayOffset(arr, elemSize=4, index=9)
\end{lstlisting}

Unfortunately, it is difficult to reliably decide on an array element size
after flattening that can be used at the time of allocation, since tuples can carry
elements that differ in size. Our tree-structured array has no information
about flattening and the access scheme generated from MLton after
flattening cannot work with our chunked array model. Hence, we need to
disable some of the flattening optimization passes. We first tried
disabling all the flattening passes including local flatten and deep
flatten. But in our later investigation, only deep flatten will try to
flatten objects in arrays. The local flatten passes are totally
compatible with our implementation. The effects of disabling /
enabling flatten passes are detailed in Section~\ref{sec:results}.

\subsection{Allocator Limitations}
{\textbf{Relative addressing for arrays:}} An example of relative
addressing is: given {\tt p = \&a[5]; } then {\tt *(p + 5)} refers to
the value stored in {\tt a[10]}. Such addressing is not used in
the SML library but is crucial to connect to C via FFI. Our
current implementation does not allow
relative addressing for arrays. This problem is extremely
common when dealing with strings, especially when passing strings
around to IO calls such as {\tt printf} since almost all IO
calls assume a continuous buffer layout. Our temporary fix is to
set an array pointer in the first leaf. However, we note that such an implementation restricts
string sizes to 128 bytes or less, which can be annoying when dealing
with file IO operations.

{\textbf{Infinite precision integers: }} MLton implements infinite
precision integers with the GMP library. MLton allocates an infinite
integer struct by first creating an object header similar to an array
header, then it maps the struct to the frontier's position and adjusts
frontier accordingly. If the calculated result can fit into a
fixed-sized representation, MLton resets frontier and converts the
representation of the integer. However, such a scheme is difficult to
realize in our model as there is no direct conversion between an
array chunk and an object chunk. We plan to look into this issue in
the future.

{\textbf{Mapping and reducing over arrays: }} Operations that span
over whole arrays are implemented in terms of array random
access in MLton's basis library. Example of {\tt Array.foldl} is
implemented in the Fig.~\ref{fig:code} (simplified, without boundary check).

\begin{figure}
\begin{lstlisting}[frame=single, basicstyle=\footnotesize\ttfamily,
  breaklines=true]
fun foldl f a arr =
  let
    val l = Array.length arr
    fun loop (i, acc) =
      if i >= l
      then acc
      else loop (i + 1, f (acc, Array.sub (arr, i)))
  in
    loop (0, a)
  end
\end{lstlisting}
\caption[]{MLton Fold Implementation.}
\label{fig:code}
\end{figure}

In MLton's representation, this implementation is fast -- accessing to
each element incurs $O(1)$ cost. But this implementation induces
unnecessary overhead in our scenario due to
$O(log(n))$ accessing time to each element. Since leaf nodes are
connected by a next pointer, we could implement these functions in
terms of MLton's FFI that maps MLton calls to C runtime. Delivering
an efficient array module requires considerable time investment and it is
left as our future work.

\subsection{GC Limitations}
\label{subsec:limit}
For collection, our GC leverages a traditional mark and sweep scheme.
We start with global objects and traverse object pointers
to mark all chunks that are live. Then we sweep through the heaps and
return the chunks that are not marked. Theoretically, a GC
is performed whenever there is insufficient memory to
allocate a new object. Yet, MLton only records
stack frames prior to current function. Objects in current stack frame are
accessed by adding the offset to {\tt stackTop}. In other words, we cannot
simply iterate through the stack to mark all objects as the layout of
the current stack frame is unknown. An alternative option is to rely
on MLton's GC checks.

MLton performs complicated data flow and control flow analysis to
insert GC checkpoints to minimize the number of garbage collections
needed. However, the data flow and control flow analysis assumes a
single heap model and objects are calculated by number of bytes
required, which is incompatible with our model. One solution is to
patch up each path in the GC check flow, redirecting all GC checks to
our GC runtime function and let the C runtime function decide
whether a garbage collection is needed. However, as will be discussed
in Section \ref{sec:results}, such a method introduces high
overhead. We currently add an RSSA optimization pass which sums up
the allocations in a block and inserts a check to see if there
are adequate chunks left. If the block does not allocate objects at
all, we ignore it. Such a check only introduces a branch and an inlined
integer comparison, which is much faster than the former method. Since arrays are allocated
in the C runtime, MLton ensures the stack is completely prepared before
jumping into {\tt GC\_arrayAllocate}. We can thus safely make GC
checks in the array allocation.

To further speed up object allocation, MLton caches the frontier in a
register. Yet, we discovered that with our chunked allocation, the
frontier is not properly flushed correctly due to changes to the control flow of
the program. We currently disable the frontier cache, but expect to be able to
infer flush points with a specialized analysis.
 MLton also uses carding to optimize the GC. Since the card map is not present
in our heap layout, we need to disable it to ensure correct
behavior.

\subsection{Potential Optimizations}
Our GC takes about 1.5x the time of MLton with corresponding
optimization passes disabled, which is far from optimal.
We are currently investigating the following to improve performance:

\begin{itemize}
\item {\textbf{CFG-based GC Checks:}} we can perform control flow
  analysis to track down the object allocation and insert GC checks at
  the start of a group of blocks and loop entries / exits instead of
  doing it per block;
\item {\textbf{Improving Object Allocation and Referencing:}} Standard ML
  provides rich type information that we may leverage to increase
  object allocation and reference efficiency, since we are able to
  figure out the number of chunks of the object and offset of the
  reference at compile time;
\item {\textbf{Re-enabling MLton's Optimization Passes:}} We will
  examine those disabled optimization passes to explore the
  possibility of making them compatible with our object layout.
\end{itemize}

\section{Porting MLton to RTEMS}
RTEMS API emulates POSIX in many places but it is not 100\%
compatible. We ported MLton to RTEMS 4.10.2 instead of the latest 4.11,
due to an issue when compiling GMP with GCC 4.9.2 provided by RTEMS
Source Builder.  MLton calls {\tt{mmap}} to allocate the heap and
{\tt{munmap}} to release it, both of which are missing as of RTEMS
4.10.2. We used {\tt malloc} and {\tt free} to manage the memory
directly.  MLton determines object alignment based on page size. Due
to lack of virtual memory, the page size for RTEMS is rather arbitrary
and we have coded it to be 1 MB. {\tt rlimit} is missing in RTEMS, and so we
needed to fall back to MLton's compatible implementation originally
intended for MinGW. RTEMS' network structure is not compatible in
various places, for example, there is no definition for {\tt
  socket\_len}. We had to strip all the POSIX networking primitives
from MLton to ensure a successful compilation.


\section{Results}
\label{sec:results}
We compare our current prototype of RT MLton with various configurations
of vanilla MLton.  Both versions of MLton
are based on MLton Git commit 2a2ebce6d12f7fc40. The evaluation is
conducted on a workstation with Intel Core i7-3770 3.4GHz CPU, running
Gentoo Linux. Fig.~\ref{fig:allresults} shows our preliminary numbers.

\begin{figure*}
  \centering
  \begin{subfigure}[b]{.33\textwidth}
    \includegraphics[width=\textwidth]{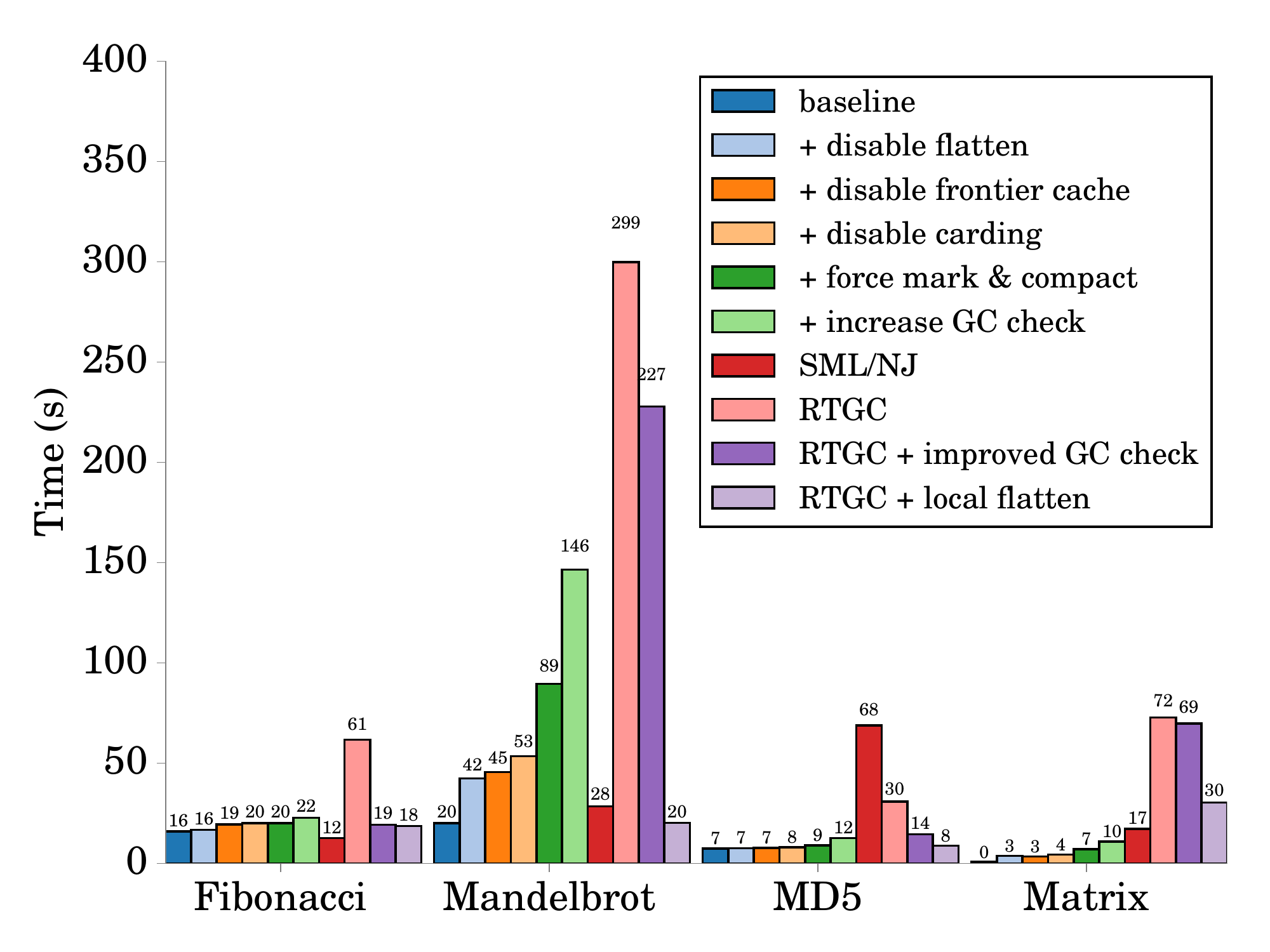}
    \caption{Baseline Performance Comparison}
    \label{fig:performance-rtgc}
  \end{subfigure}
  \begin{subfigure}[b]{.33\textwidth}
    \centering
    \includegraphics[width=\textwidth]{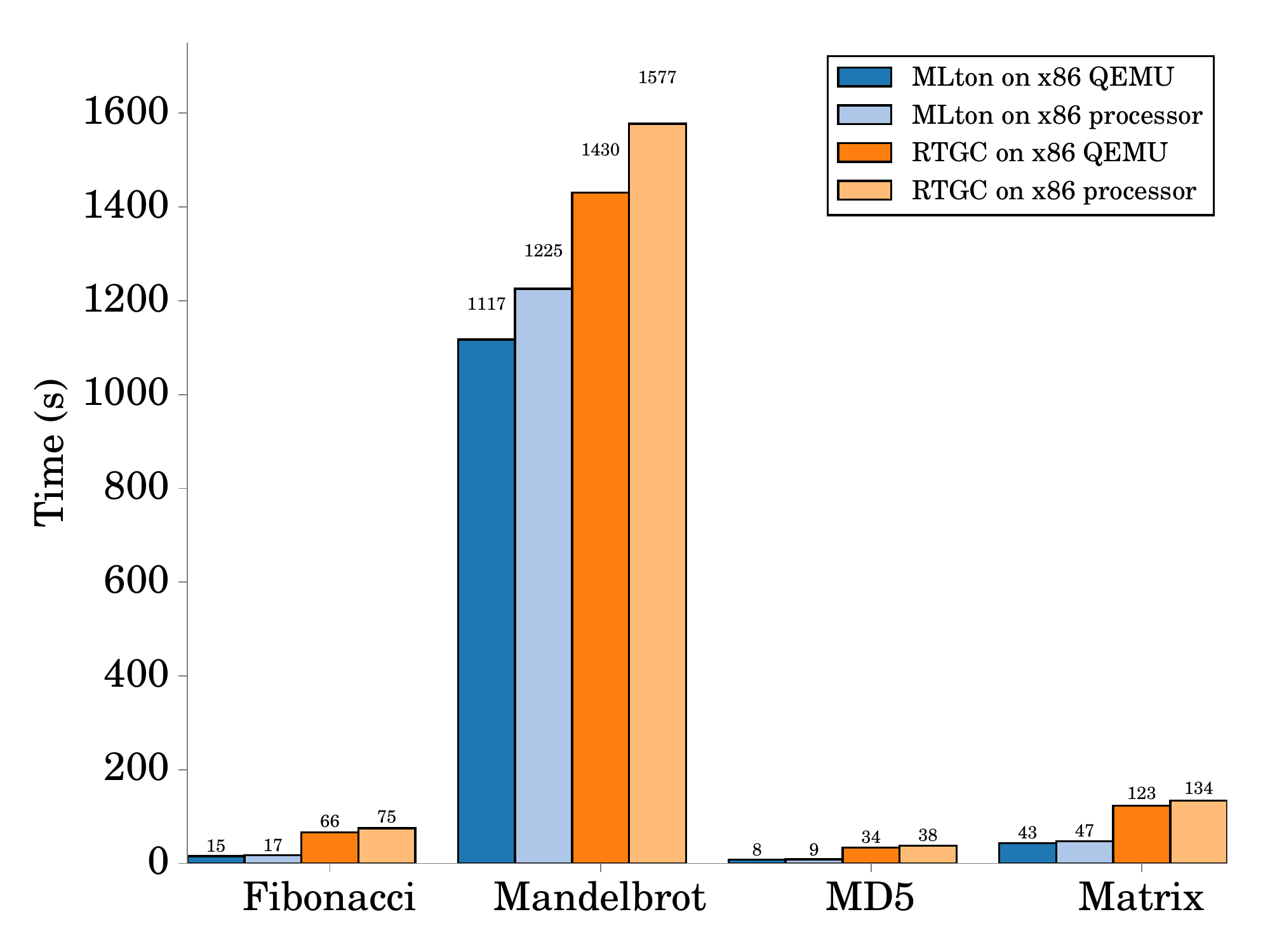}
    \caption{Embedded Performance on RTEMS}
    \label{fig:rtems-performance}
  \end{subfigure}
  \begin{subfigure}[b]{.33\textwidth}
    \includegraphics[width=\textwidth]{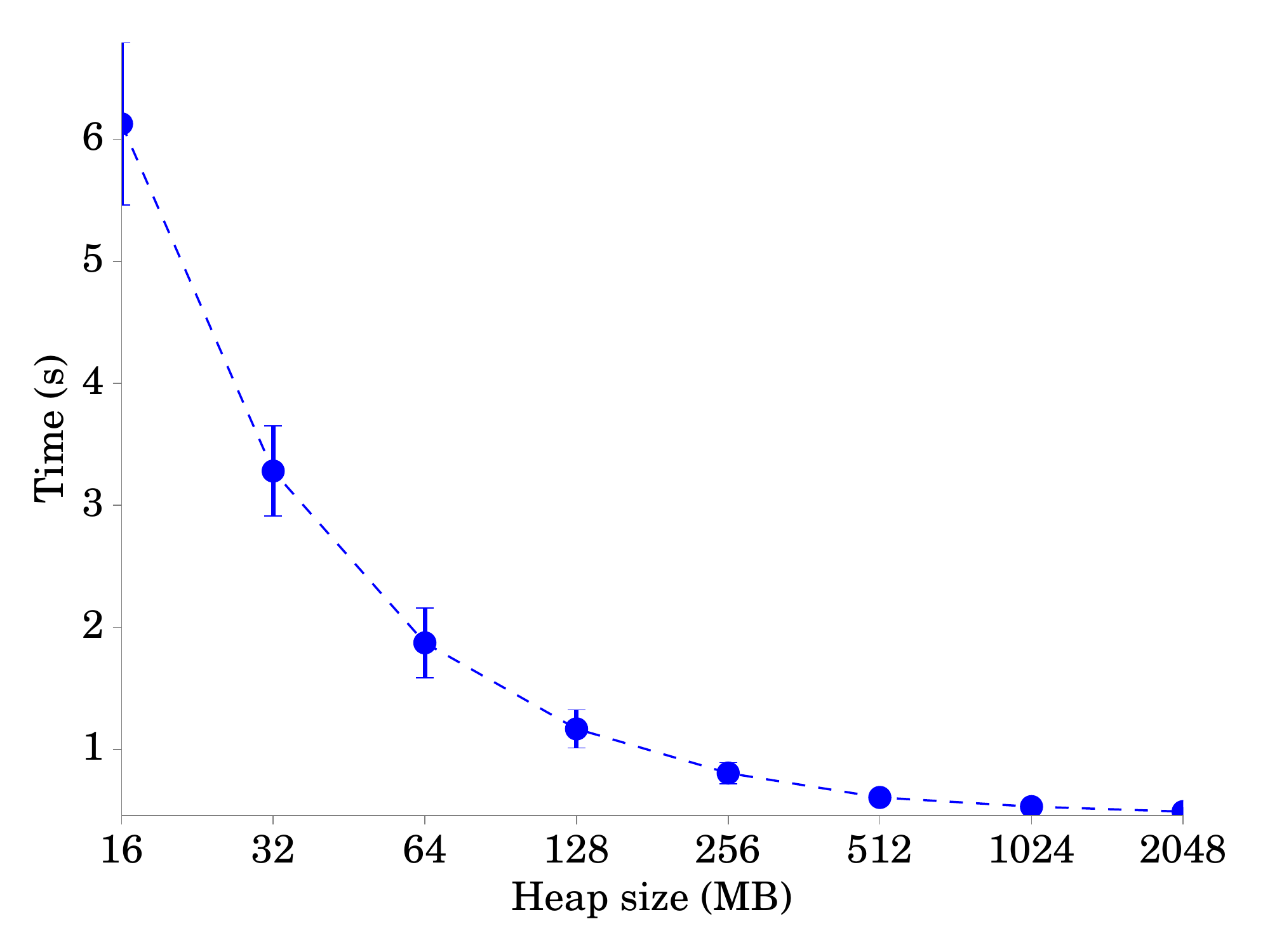}
    \caption{RTGC time vs heap size}
    \label{fig:heap-size-gc}
  \end{subfigure}
  \caption{Performance metrics of RT MLton}
  \label{fig:allresults}
\end{figure*}

MLton makes exceptional effort to optimize the final
program. Unfortunately, we need to disable various optimization passes
for correctness as mentioned in Section~\ref{sec:hybrid-gc}. In
Fig.~\ref{fig:performance-rtgc}, we demonstrate the effect of
disabling optimization passes in MLton and compare our implementation
with those configurations. Disabling optimization passes increases
code size and slows the performance for MLton by multiple
times. Compared to MLton with those optimizations disabled, our
current implementation is approximately 3x slower. We are unfortunately
very conservative at inserting GC checks at each path, resulting in
interrupted control flow through jumps to
a (potentially unnecessary) GC checkpoint. This also prevents further optimization passes from taking
effect. Table.~\ref{tbl:gc-check-code-size} compares the code size
before / after limit check and at the start of machine IL (right after
RSSA finished). We add an RSSA optimization pass to insert GC checks
block-wise. As shown in Fig.~\ref{fig:performance-rtgc}, it improves
the performance dramatically. In the Fibonacci and MD5 benchmark, our
implementation achieves identical performance to MLton with
optimization passes disabled. Array referencing is particularly slow at
the moment as demonstrated in the Matrix Multiplication
benchmark. Currently, to allocate and iterate through an array of
10,000,000 integers, our array implementation takes 0.3 seconds. The
Matrix Multiplication benchmark multiplies matrices heavily based on
individual element reference that introduces $O(log(n))$ cost at each
access. We believe it could be sped up
dramatically if the program is made aware of our architecture and
implements matrix multiplications as reduction over continuous
sequences and with the delivery of sequential access optimizations
discussed in Section~\ref{subsec:limit}. We envision that to ensure
performance in real time setting, a functional program still needs
to be aware of the low level object model and optimize algorithm /
code towards the model. Fig.~\ref{fig:performance-rtgc}
also shows the potential benefit of enabling the local flattening
passes while leaving deep flattening off. Local flattening ensures
smaller object sizes and significantly reduces allocations. In fact,
in Mandelbrot and MD5, with local flattening enabled, the GC is never
activated. The result in this case is very similar to the vanilla
MLton, which proves that the chunked object access adds only little
overhead.

\begin{table}
  \begin{tabular}{ l | c | c }
    & Vanilla MLton & Conservative limit check \\\hline
    Before limit check & 57,328 bytes & 54,956 bytes \\
    After limit check & 77,904 bytes & 179,324 bytes \\
    Machine IL & 103,028 bytes & 216,960 bytes
  \end{tabular}
  \caption{Impact of limit check on RSSA optimization}
  \label{tbl:gc-check-code-size}
\end{table}

To evaluate the actual performance on a real time system, we have
conducted evaluation on RTEMS both in x86 QEMU and real x86 CPU. The
workstation has Intel Xeon E3-1230 v2 3.3GHz. QEMU is configured with
4G available memory and KVM enabled. Direct execution on the CPU is
enabled by connecting GRUB with the compiled RTEMS
executable. Fig.~\ref{fig:rtems-performance} depicts the results. We
observe a smaller performance gap between our implementation and
vanilla MLton. In both cases, our implementation is at 3 - 5x slower
than vanilla MLton. Directly running RTEMS executable on the CPU
results in a slightly slower performance, which may due to Linux
Kernel's awareness of Turbo Boost of the CPU. We haven't observed as
significant performance boost as running on x86 Linux by refining the
limit checks -- generally, it runs only 1 - 2 s faster than the
numbers in Fig.~\ref{fig:rtems-performance}. Note that we make every
effort to ensure the comparison is fair but various factors could
contribute to the performance results, including of platform specific
functions to RTEMS largely based on MinGW platform implementation for
MLton, which might not be efficient when used on RTEMS. MLton's lack
of knowledge of RTEMS may be a contributing factor to the
ineffectiveness of its optimizations. Old version (4.4) of GCC that
lacks particular knowledge of certain flow optimization used by RTEMS
4.10 tool-chain might cause inefficient code organization. Targeting
i386 instead of i686 in our RTEMS port may disable a few
optimizations. We thus consider this result as a preliminary baseline,
but expect better performance once we consider optimizations geared
toward RTEMS and embedded execution.

Fig.~\ref{fig:heap-size-gc} depicts the relation between time taken by garbage
collection for our RTGC compared to heap size while running Mandelbrot set benchmark of size
8192.



\section{Related Work}
\label{sec:related}

{\textbf{Real-Time Garbage Collection:}}
There are roughly three classes of RTGC:
(i) {\em time based}~\cite{Bacon:2003:CFS:780732.780744} where the GC is scheduled as a task in the system,  (ii) {\em slack
based}~\cite{fijiGC} where the GC is the lowest priority real-time task and executes in the times between release of higher
priority tasks, and (iii) {\em work based}~\cite{Siebert:2007:RGC:1288940.1288954} where each allocation triggers an amount of GC work proportional to the allocation request.
 In each of these RTGC definitions, the overall system designer {\em must} take into consideration the time requirements to run the RTGC. 
 We currently have adopted a slack based approach in the context of real-time MLton, though a work based approach is also worth
 exploring.

{\textbf{Real-Time Java:}} The real-time specification for Java RTSJ~\cite{RTSJ} and safety critical Java (SCJ)~\cite{scj}, both provide
definitions for scoped memory~\cite{Hamza:2012:RRM}, a region based automatic memory management scheme. We believe it would be interesting to leverage previous
work on region inference~\cite{Deters:2002} in the context of ML~\cite{Tofte:1997:RMM:249657.249661} to eschew RTGC entirely through the
use of scoped memory.

\section{Conclusion and Future Work}
\label{sec:conc}
In this paper we presented our prototype implementation of a real-time capable version of MLton.  Our next
steps are to investigate necessary changes to the optimization passes we needed to disable to ensure correctness
and predictability.  We also plan on investigating new optimizations, specifically targeted at reducing the overheads
of the hybrid fragmenting GC.  Previous experience with the Fiji VM indicates that this is feasible.  Lastly, we will consider
optimizations specific to RTEMS and revisit the I/O libraries.

\balance
\bibliographystyle{abbrv}
\small
\bibliography{all}

\end{document}